%% file: ms.tex
\documentclass[letterpaper,twocolumn,10pt]{article}
\usepackage{usenix,epsfig}
\usepackage{epsf}
\usepackage{epsfig}
\usepackage{graphicx}
\usepackage{times}
\usepackage{ifthen}
\usepackage{comment}
\usepackage[hyphens]{url}
\usepackage{hyperref}
\usepackage{latexsym}
\usepackage[margin=1in]{geometry}
\usepackage{amsfonts}
\usepackage{subfigure}
\usepackage{amsmath}
\usepackage{multirow}
\usepackage{floatflt}
\usepackage{listings}
\usepackage[dvipsnames]{xcolor}
\usepackage{xspace}
\usepackage{enumitem}
\usepackage[compact]{titlesec}
\usepackage{mathpazo,euler}
\usepackage{amssymb}
\usepackage{ifthen}
\usepackage{tabularx}
\usepackage{makecell}
\usepackage{tikz}
\include{macro}

\definecolor{dkgreen}{rgb}{0,0.6,0}
\definecolor{gray}{rgb}{0.5,0.5,0.5}
\definecolor{mauve}{rgb}{0.58,0,0.82}
\definecolor{shadecolor}{rgb}{0.95,0.95,0.95}
\begin{document}

\date{}

\title{\Large \bf Discovering Flaws in Security-Focused Static Analysis Tools for Android using Systematic Mutation}

\author{
{\rm Richard Bonett, Kaushal Kafle, Kevin Moran, Adwait Nadkarni, Denys
Poshyvanyk}\\
\{rfbonett, kkafle, kpmoran, nadkarni, denys\}@cs.wm.edu\\
William \& Mary
}

\maketitle

\thispagestyle{empty}

\input{abstract}

\input{intro}

\input{motivation}

\input{overview}

\input{design}

\input{implementation}

\input{evaluation}

\input{discussion}

\input{related-work}

\input{limitations}

\input{conclusion}

\input{acknowledgements}

{\footnotesize \bibliographystyle{acm}
\bibliography{bib/ms.bib}}

\input{appendix}

\end{document}

%% file: macro.tex
\newcommand{\tool}{$\mu$SE\xspace}
\newcommand{\tools}{$\mu$SE's\xspace}

\newboolean{showcomments}

\setboolean{showcomments}{true}

\ifthenelse{\boolean{showcomments}}
  {\newcommand{\nb}[2]{
    \fbox{\bfseries\sffamily\scriptsize#1}
    {\sf\small$\blacktriangleright$\textit{#2}$\blacktriangleleft$}
   }
   
  }
  {\newcommand{\nb}[2]{}
   
  }

\newcommand\myparagraph[1]{\noindent {\bf {#1}:}}

\newcommand{\ie}{\textit{i.e.,}\xspace}
\newcommand{\eg}{\textit{e.g.,}\xspace}

\newcommand{\etal}{\textit{et al.}\xspace}
\newcommand{\etals}{\textit{et al.'s}\xspace}


%% file: abstract.tex
\begin{abstract}

Mobile application security has been one of the major areas of security
research in the last decade. Numerous application analysis tools have
been proposed in response to malicious, curious, or vulnerable
apps.
However, existing tools, and specifically, static analysis tools, trade
soundness of the analysis for precision and performance, and are hence
sound{\em y}. Unfortunately, the specific unsound choices or flaws in
the design of these tools are often not known or well-documented,
leading to a misplaced confidence among researchers, developers, and
users.
This paper proposes the {\em Mutation-based soundness evaluation}
(\tool) framework, which systematically evaluates Android static
analysis tools to discover, document, and fix, flaws, by
leveraging the well-founded practice of mutation analysis.
We implement \tool as a semi-automated framework, and apply it to a set
of prominent Android static analysis tools that detect private data
leaks in apps.
As the result of an in-depth analysis of one of the major tools, we
discover $13$ undocumented flaws. More importantly, we discover that all
$13$ flaws propagate to tools that inherit the flawed tool. We
successfully fix one of the flaws in cooperation with the tool
developers. 
Our results motivate the urgent need for systematic discovery
and documentation of unsound choices in sound{\em y} tools, and
demonstrate the opportunities in leveraging mutation testing in achieving this goal.

\end{abstract}

%% file: intro.tex
\section{Introduction}
\label{sec:intro}

Mobile devices such as smartphones and tablets have become the fabric of
our  consumer computing ecosystem; by the year $2020$, more than $80$\% of
the world's adult population is projected to own a
smartphone~\cite{economist}.  
This popularity of mobile devices is driven by the millions of diverse,
feature-rich, third-party applications or ``apps'' they support.
However, in fulfilling their functionality, apps often require access to
security and privacy-sensitive resources on the device (\eg  GPS
location, security settings). Applications can neither be trusted to be
well-written or benign, and to prevent misuse of such access through
malicious or vulnerable
apps~\cite{lnw+14,gzz+12,zj12,rce13,fhm+12,ssl+14,ebfk13}, it is
imperative to understand the challenges in securing mobile apps.

Security analysis of third-party apps has been one of the dominant areas
of smartphone security research in the last decade, resulting in tools
and frameworks with diverse security goals. For instance, prior work has
designed tools to identify malicious behavior in
apps~\cite{eom09b,zwzj12,ash+14}, discover private data
leaks~\cite{egc+10,arf+14,gcec12,akg+15}, detect vulnerable application
interfaces~\cite{fwm+11,cfgw11,llw+12,lbs+17}, identify flaws in the use
of cryptographic primitives~\cite{fhm+12,ebfk13,ssl+14},
and define sandbox policies for third-party apps~\cite{hdge14,jsz16}.  
To protect users from malicious or vulnerable apps, it is imperative to
assess the challenges and pitfalls of existing tools and techniques.
However, {\em it is unclear if existing security tools are robust enough
to expose particularly well-hidden unwanted behaviors.}

Our work is motivated by the pressing need to discover the limitations
of application analysis techniques for Android. Existing application
analysis techniques, specifically those that employ static analysis,
must in practice trade soundness for precision, as there is an
inherent conflict between the two properties. A sound analysis requires
the technique to over-approximate (\ie consider instances of unwanted
behavior that may not execute in reality), which in turn deteriorates
precision. This trade-off has practical implications on the security
provided by static analysis tools.  That is, {\em in theory}, static
analysis is expected to be sound, yet, in practice, these tools must
purposefully make unsound choices to achieve a feasible analysis that has
sufficient precision and performance to scale. For instance, techniques
that analyze Java generally do not over-approximate analysis of certain
programming language features, such as reflection, for practical reasons
(\eg Soot~\cite{vcg+99}, FlowDroid~\cite{arf+14}). While this particular
case is well-known and documented, many such unsound design choices are
neither well-documented, nor known to researchers outside a small
community of experts.

Security experts have described such tools as sound{\em y}, \ie having a
core set of sound design choices, in addition to certain practical
assumptions that sacrifice soundness for precision~\cite{lss+15}. While
soundness is an elusive ideal, sound{\em y} tools certainly seem to be
a practical choice: {\em but only if the unsound choices are known,
necessary, and clearly documented}. However, the present state of
sound{\em y} static analysis techniques is dire, as unsound choices {\sf
(1)}~may not be documented, and unknown to non-experts, {\sf (2)}~may
not even be known to tool designers (\ie implicit assumptions), and {\sf
(3)}~may propagate to future research.  The sound{\em i}ness manifesto
describes the misplaced confidence generated by the insufficient
study and documentation of sound{\em y} tools, in the specific context
of language features~\cite{lss+15}. While our work is motivated by the
manifesto, we leverage sound{\em i}ness at the general, conceptual level
of design choices, and attempt to resolve the status quo of sound{\em y}
tools by making them more secure as well as transparent.

This paper proposes the  {\em Mutation-based Soundness Evaluation}
(\tool, read as ``muse'') framework that enables systematic security
evaluation of Android static analysis tools to discover unsound
design assumptions, leading to their documentation, as well as
improvements in the tools themselves. \tool leverages the practice of
mutation analysis from the software engineering (SE)
domain~\cite{Offutt2001,Hamlet:TSE,DeMillo:Computer,Ma:ISSRE03,Derezinska2014,Praphamontripong:Mutation15,Appelt:ISSTA14,ZhouF09,Oliveira:ICSTW15,Nardo:ICST15},
and specifically, more recent advancements in mutating Android
apps~\cite{lbt+17}. 
In doing so, \tool adapts a well-founded practice from SE to security, by
making useful changes to contextualize it to evaluate security tools.

\tool creates {\em security operators}, which reflect the security goals
of the tools being analyzed (\eg data leak or SSL
vulnerability detection). These security operators are seeded, \ie
inserted into one or more Android apps, as guided by a {\em
mutation scheme}. This seeding results in the creation of multiple
mutants (\ie code that represents the target unwanted behavior) within
the app. Finally, the mutated application is analyzed
using the security tool being evaluated, and the undetected mutants are
then subjected to a deeper analysis. We propose a semi-automated methodology to
analyze the uncaught mutants, resolve them to flaws in the
tool, and confirm the flaws experimentally. 

We demonstrate the effectiveness of \tool by evaluating static analysis
research tools that detect data leaks in Android apps (\eg
FlowDroid~\cite{arf+14}, IccTA~\cite{lbb+15}). We evaluate a set of
$seven$ tools across three experiments, and reveal $13$ flaws
that were undocumented. We also discover that when a tool inherits
another (\ie inherits the codebase), {\em all the flaws propagate}. Even
in cases wherein a tool only conceptually inherits another (\ie
leveraging decisions from prior work), {\em just less than half of the
flaws propagate}. We provide immediate patches that fix one
flaw, and in other cases, we identify flaw classes that may need
significant research effort.  Thus, \tool not only helps researchers,
tool designers, and analysts uncover undocumented flaws and
unsound choices in sound{\em y} security tools, but may also provide
immediate benefits by discovering easily fixable, but evasive,
flaws.

This paper makes the following contributions:\vspace{-0.5em}
\begin{itemize}\renewcommand{\itemsep}{-0.1em}
  \item {\em We introduce the novel paradigm of Mutation-based Soundness
    Evaluation}, which provides a systematic methodology for discovering
    flaws in static analysis tools for Android, leveraging the
    well-understood practice of mutation analysis.  We adapt mutation
    analysis for security evaluation, and design the abstractions of {\em security
    operators} and {\em mutation schemes}.
  \item {\em We design and implement the \tool framework} for evaluating
    Android static analysis tools. \tool adapts to the security goals of
    a tool being evaluated, and allows the detection of unknown or undocumented flaws. 
  \item {\em We demonstrate the effectiveness of \tool by evaluating
    several widely-used Android security tools} that detect
    private data leaks in Android apps.  \tool detects $13$ unknown
    flaws, and validates their propagation. Our analysis
    leads to the documentation of unsound assumptions, and immediate
    security fixes in some cases.
\end{itemize}

\myparagraph{Threat Model}
\tool is designed to help security researchers evaluate tools that
detect vulnerabilities (\eg SSl misuse), {\em and more
importantly}, tools that detect malicious or suspicious behavior (\eg
data leaks). Thus, the security operators and mutation schemes defined
in this paper are of an adversarial nature. That is, behavior like
``data leaks'' is intentionally malicious/curious, and generally not
attributed to accidental vulnerabilities. Therefore, to evaluate the
soundness of existing tools that detect such behavior, \tool has to
develop mutants that mimic such adversarial behavior as well, by
defining mutation schemes of an adversarial nature.  This is the key
difference between \tool and prior work on fault/vulnerability injection
(e.g., LAVA~\cite{dhk+16}) that assumes the mutated program to be
benign.

The rest of the paper proceeds as follows:
Section~\ref{sec:motivation} motivates our approach, and
provides a brief background. Section~\ref{sec:overview} describes the
general approach and the design goals. Section~\ref{sec:design}
and Section~\ref{sec:implementation} describe the design and
implementation of \tool, respectively.
Section~\ref{sec:eval} evaluates the
effectiveness of \tool, and Section~\ref{sec:discussion}
delivers the insights distilled from it. Section~\ref{sec:relwork}
describes related work. Section~\ref{sec:limitations} describes
limitations. Section~\ref{sec:conc} concludes.

%% file: motivation.tex
\vspace{-0.5em}
\section{Motivation and Background}
\label{sec:motivation}
\vspace{-0.5em}

This work is motivated by the pressing need to help
researchers and practitioners identify instances of unsound
assumptions or design decisions in their static analysis tools, thereby {\em
extending the sound core} of their sound{\em y} techniques.  That is,
security tools may already have a core set of sound design decisions (\ie
the sound core), and may claim soundness  based on those decisions.
While the soundiness manifesto~\cite{lss+15} defines the {\em sound
core} in terms of specific language features, we use the term in a more
abstract manner to refer to the design goals of the tool.
Systematically identifying unsound decisions may allow researchers to
resolve flaws and help extend the sound
core of their tools.

Moreover, research papers and tool documentations indeed do not
articulate many of the unsound assumptions and design choices that lie
outside their sound core, aside from some well-known cases (\eg choosing
not to handle reflection, race conditions), as confirmed by our results
(Section~\ref{sec:eval}). There is also a chance that developers of
these techniques may be unaware of some implicit assumptions/flaws due
to a host of reasons: \eg  because the assumption was inherited from
prior research or a certain aspect of Android was not modeled correctly.
Therefore, our objective is to discover instances of such hidden
assumptions and design flaws that affect the security claims made by
tools, document them explicitly, and possibly, help developers and
researchers mend existing artifacts.

\subsection{Motivating Example}
\label{sec:example}
\vspace{-0.5em}
Consider the following motivating example of a prominent static analysis tool,
FlowDroid~\cite{arf+14}:

FlowDroid~\cite{arf+14} is a highly popular static analysis framework for
detecting private data leaks in Android apps by performing a data flow
analysis.  Some of the limitations of FlowDroid are well-known and
stated in the paper ~\cite{arf+14}; \eg FlowDroid does not support
reflection, like most static analyses for Java.
However, through a systematic evaluation of FlowDroid, we
discovered a security limitation that is not well-known or accounted for
in the paper, and hence affects guarantees provided by the tool's analysis.  
We discovered that FlowDroid (\ie v1.5, latest as of 10/10/17) does not support ``Android
fragments''~\cite{android-fragments}, which are app modules that are
widely used in most Android apps (\ie in more than 90\% of the top 240
Android apps per category on Google Play, see Appendix A). This flaw renders any security
analysis of general Android apps using FlowDroid unsound, due to the
high likelihood of fragment use, even when the app developers may be
cooperative and non-malicious.  Further, FlowDroid v2.0, which was recently
released~\cite{fdroid_release}, claims to address fragments, {\em but
also failed to detect our exploit}.  
On investigating
further, we found that FlowDroid v1.5 has been extended by at least 13
research tools~\cite{lbb+15, kfb+14, Yang:2015, akg+15, Octeau:2015,
Sasnauskas:2014, Liu:2015, Slavin:2016, Aafer:2015, Rasthofer:2014,
Li:2014, Lillack:2017, Nan:2015},
none of which acknowledge or address this
limitation in modeling fragments. This leads us
to conclude that this significant flaw not only persists in
FlowDroid, but may have also propagated to the tools that inherit it.
We confirm this conjecture for inheritors of FlowDroid
that also detect data leaks, and are available in source or binary form
(\ie $2$ out of $13$), in Section~\ref{sec:eval}.

Finally, we reported the flaws to the authors of FlowDroid,
and created two patches to fix it.
Our patches were
confirmed to work on FlowDroid v$2.0$ built from source, and were
accepted into FlowDroid's repository~\cite{appendix}.  Thus, we were
able to discover and fix an undocumented design flaw that
significantly affected FlowDroid's soundness claims, thereby expanding
its sound core.  However, we have confirmed that
FlowDroid v$2.5$~\cite{fdroid_release} still
fails to detect leaks in fragments, and are working with developers
to resolve this issue.

Through this example, we demonstrate that unsound assumptions in
security-focused static analysis tools for Android are not only
detrimental to the validity of their own analysis, but may also
inadvertently propagate to future research.  Thus, identifying these
unsound assumptions is not only beneficial for making the user of the
analysis aware of its true limits, but also for the research community
in general.  As of today, aside from a handful of manually curated
testing toolkits (\eg DroidBench~\cite{arf+14}) with hard-coded (but
useful) checks, to the best of our knowledge, there has been no prior
effort at methodologically discovering problems related to sound{\em
i}ness in Android static analysis tools and frameworks.  {\em This paper
is motivated by the need to systematically identify and resolve the unsound
assumptions in security-related static analysis tools.} 

\vspace{-0.4em}
\subsection{Background on Mutation Analysis}
\label{subsec:background}
\vspace{-0.2em}

Mutation analysis has a strong foundation in the field of SE, and is
typically used  as a test adequacy criterion, measuring the
effectiveness of a set of test cases~\cite{Offutt2001}. Faulty programs
are created by applying transformation rules, called \textit{mutation
operators} to a given program. The larger the number of faulty
programs or \textit{mutants} detected by a test suite, the higher the
effectiveness of that particular suite. 
Since its inception~\cite{Hamlet:TSE,DeMillo:Computer}, mutation testing has seen striking advancements related to
the design and development of advanced operators. Research related to
development of mutation operators has traditionally attempted to adapt
operators for a particular target domain, such as the web~\cite{Praphamontripong:Mutation15}, data-heavy applications
\cite{Appelt:ISSTA14,ZhouF09,Nardo:ICST15}, or GUI-centric applications
\cite{Oliveira:ICSTW15}.  
Recently, mutation analysis has been applied to
measure the effectiveness of test suites for both functional and
non-functional requirements of Android apps
\cite{Deng:ICSTW15,Jabbarvand:FSE17, lbt+17}.

This paper builds upon SE concepts of mutation analysis
and adapts them to a security context. Our methodology
does not simply use the traditional mutation analysis, but
rather {\em redefines} this methodology to effectively improve
security-focused static analysis tools, as we describe in
Sections~\ref{sec:design} and ~\ref{sec:relwork}.

%% file: overview.tex
\section{\tool}
\label{sec:overview}

\begin{figure}[t]
    \centering
    \includegraphics[width=3in]{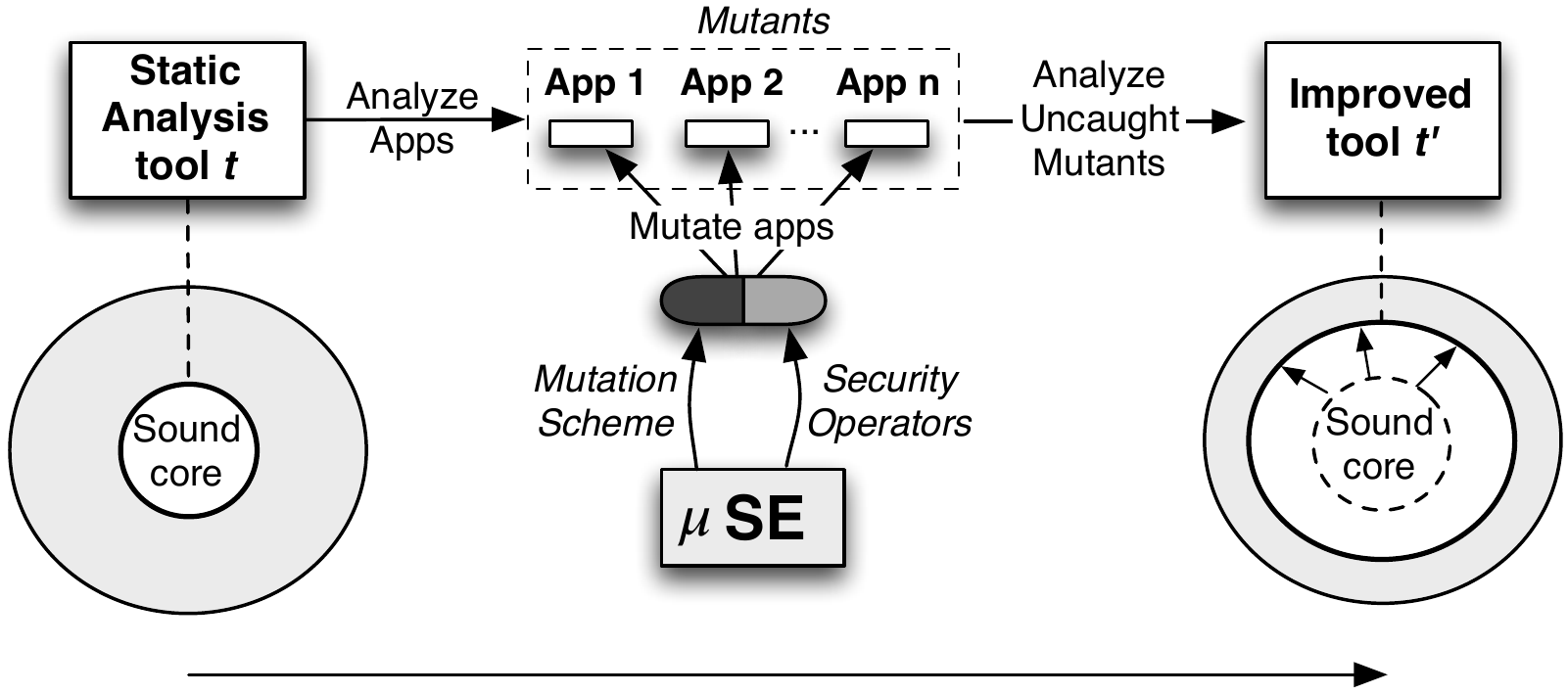} 
    \vspace{-1em}
    \caption{{\small \tool tests a static analysis tool on a set of
    mutated Android apps and analyzes uncaught mutants to discover and/or
    fix flaws.}}
    \vspace{-1.5em}
\label{fig:expanding_core}
\end{figure} 

We propose \tool, a semi-automated framework for systematically
evaluating Android static analysis tools that adapts the process of
mutation analysis commonly used to evaluate software test
suites~\cite{Offutt2001}. That is, we aim to help discover concrete
instances of flawed security design decisions made by static analysis
tools, by exposing them to methodologically mutated applications. We
envision two primary benefits from \tool: {\em short-term} benefits
related to straightforwardly fixable flaws that may be patched
immediately, and  {\em long-term} benefits related to the continuous
documentation of assumptions and flaws, even those that may be hard to
resolve. This section provides an overview of \tool
(Figure~\ref{fig:expanding_core}) and its design goals.

 As shown in Figure~\ref{fig:expanding_core}, we take
an Android static analysis tool to be evaluated (\eg FlowDroid~\cite{arf+14} or
MalloDroid~\cite{fhm+12}) as input. \tool
executes the tool on {\em mutants}, \ie apps to which
{\em security operators} (\ie security-related mutation operators) are
applied, as per a \textit{mutation scheme}, which governs the
placement of code transformations described by operators in the app (\ie thus generating mutants). The security operators represent anomalies that the static analysis tools
are expected to detect, and hence, are closely tied to the security goal
of the tool. The uncaught mutants indicate flaws in the tool, and
analyzing them leads to the broader discovery and awareness of the
unsound assumptions of the tools, eventually facilitating
security-improvements.

\input{design-goals}

%% file: design-goals.tex

\myparagraph{Design Goals}
Measuring the security provided by a system is a difficult problem;
however, we may be able to better predict failures if the assumptions
made by the system are known in advance.  Similarly, while soundness may
be a distant ideal for security tools, we assert that it should be
feasible to articulate the boundaries of a tool's sound core. Knowing
these boundaries would be immensely useful for analysts who use security
tools, for developers looking for ways to improve tools, as well as for
end users who benefit from the security analyses provided by such tools.
To this end, we design \tool to provide an effective foundation for
evaluating Android security
tools. Our design of \tool is guided by the following
goals:\vspace{-0.5em}
\begin{enumerate}[label=\textbf{G\arabic*},ref=\textbf{G\arabic*}]
\setlength{\itemsep}{-3pt} 
  \item \label{goal:operator} {\em Contextualized security operators.}
    Android security tools have diverse purposes and may claim various
    security guarantees. Security operators must be instantiated in a
    way that is sensitive to the context or purpose (\eg data leak
    identification) of the tool being evaluated.
  \item \label{goal:android} {\em  Android-focused mutation
    scheme.} Android's security challenges are notably unique, and hence
    require a diverse array of novel security analyses.  Thus, the
    mutation schemes, \ie the {\em placement} of the target, unwanted
    behavior in the app, must consider Android's abstractions and
    application model for effectiveness.
  \item \label{goal:feasibility} {\em Minimize manual-effort during
    analysis.} While \tool is certainly more feasible than
    manual analysis, 
    we intend to  significantly reduce the manual effort spent on
    evaluating undetected mutants. Thus, our goal is to dynamically filter
    inconsequential mutants, as well as to develop a systematic
    methodology for resolving undetected mutants to flaws.

  \item \label{goal:performance} {\em Minimize overhead.} We expect
    \tool to be used by security researchers as well as tool users and
    developers. Hence, we must ensure that \tool is efficient so as to to promote a wide-scale deployment and community-based use of
    \tool.  
\end{enumerate}

%% file: design.tex
\vspace{-0.3em}
\section{Design} 
\label{sec:design}
\vspace{-0.2em}

\begin{figure}[t]
    \centering
    \includegraphics[width=2.8in]{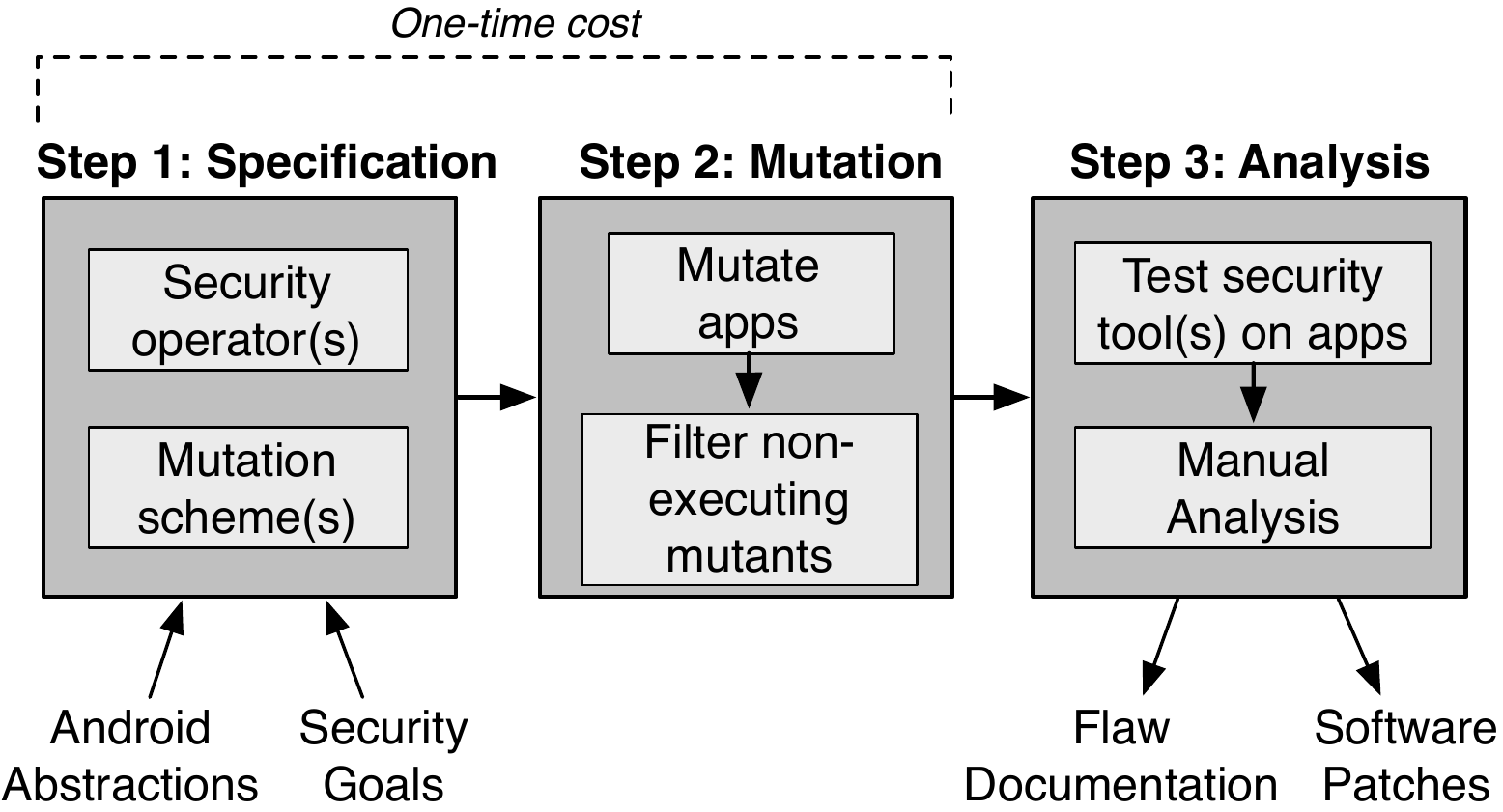} 
    \vspace{-0.5em}
    \caption{{\small The components and process of the \tool.}} 
    \vspace{-1.5em}
\label{fig:design}
\end{figure}

Figure~\ref{fig:design} provides a conceptual description of the process
followed by \tool, which consists of three main steps. In Step 1, we
{\em specify} the security operators and mutation schemes that are
relevant to the security goals of the tool being evaluated (\eg data
leak detection), as well as certain unique abstractions of Android that
separately motivate this analysis.  In Step 2, we {\em mutate} one or
more Android apps using the security operators and defined
mutation schemes using a {\em Mutation Engine (ME)}. After this step each app is
said to contain one or more mutants. To maximize effectiveness, mutation
schemes in \tool stipulate that mutants should be systematically
injected into all potential locations in code where operators can be
instantiated. In order to limit the effort required for manual analysis
due to potentially large numbers of mutants, we first filter out the
non-executing mutants in the Android app(s) using a dynamic {\em Execution
Engine (EE)} (Section~\ref{sec:implementation}). In
Step 3, we apply the security tool under investigation to {\em analyze}
the mutated app, leading it to detect some or all of the mutants as
anomalies. We perform a methodological manual analysis of the undetected
mutants, which may lead to documentation of flaws, and
software patches. 

Note that tools sharing a security goal (\eg FlowDroid\cite{arf+14}, Argus~\cite{wror14},
HornDroid~\cite{cgm16} and BlueSeal~\cite{svt+14} all detect data leaks) can be analyzed using the same
security operators and mutation schemes, and hence the mutated apps,
significantly reducing the overall cost of operating \tool
(Goal~\ref{goal:performance}).  The rest of
this section describes the design contributions of \tool. The precise
implementation details can be found in Section~\ref{sec:implementation}.

\input{security_operators}
\input{mutation_schemes}
\input{analysis_feasibility}

%% file: security_operators.tex
\subsection{Security Operators}
\label{sec:operators}

A security operator is a description of the unwanted behavior that the
security tool being analyzed aims to detect. When designing security
operators, we are faced with an important question: {\em what do we want
to express?} Specifically, the operator might be too coarse or
fine-grained; finding the correct granularity is the key. 

For instance, defining operators specific to the implementations of
individual tools may not be scalable. On the contrary, defining a
generic security operator for all the tools may be too simplistic to be effective. Consider
the following example:

\begin{figure}[t]
  \centering
  \includegraphics[width=3in]{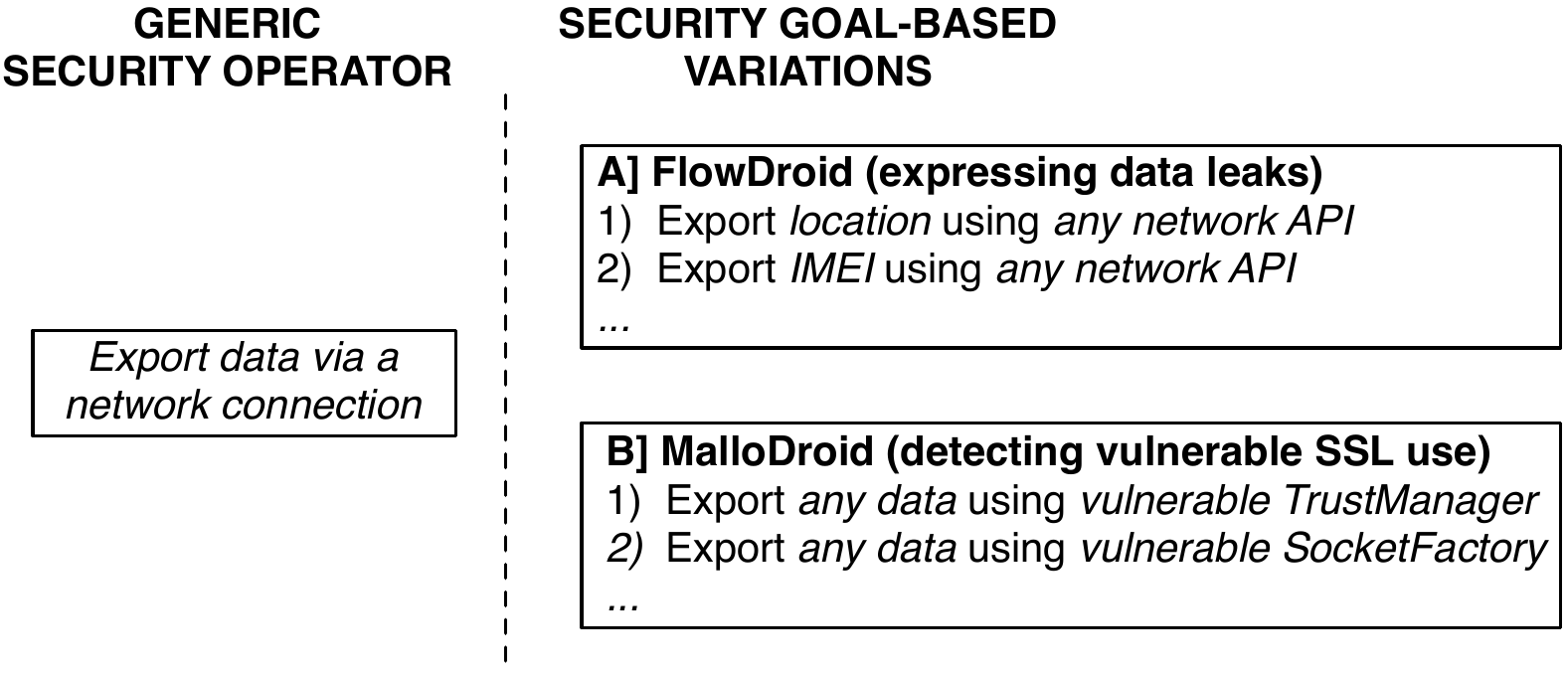}
  \vspace{-1em}
  \caption{{\small A generic "network export" security operator, and its
  more fine-grained instantiations in the context of
  FlowDroid~\cite{arf+14} and MalloDroid~\cite{fhm+12}.}}
  \vspace{-1em}
  \label{fig:instantiation}
\end{figure}

Figure~\ref{fig:instantiation} describes the limitation of using a
generic security operator that describes code which ``exports data to the
network''. Depending on the tool being evaluated, we may need a unique,
fine-grained, specification of this operator. For example, for
evaluating FlowDroid~\cite{arf+14}, we may need to express the specific
types of private data that can be exported via any of the network APIs,
\ie the data portion of the operator is more important than what
network API is used. However, for evaluating a
tool that detects vulnerable SSL connections (\eg CryptoLint~\cite{ebfk13}),
we may want to express the use of vulnerable SSL APIs (\ie of SSL
classes that can be overridden, such as a custom TrustManager that
trusts all certificates) without much concern for what data is exported.
That is, the requirements are practically orthogonal for these two use
cases, rendering a generic operator useless, while precisely designing
tool-specific operators may not scale.

In \tool, we take a balanced approach to solve this problem: instead of
tying a security operator to a specific tool, we define it in terms of
the {\em security goal} of the concerned tool (Goal~\ref{goal:operator}). Since the security goal
influences the properties exhibited by a security analysis, security
operators designed with a particular goal in consideration
would apply to all the tools that claim to have that security goal,
hence making them feasible and scalable to design. For instance, a security operator that
reads information from a private source (\eg IMEI, location)
and exports it to a public sink (\eg the device log, storage) would be appropriate to use for all the tools that
claim to detect private data leaks (\eg Argus~\cite{wror14},
HornDroid~\cite{cgm16}, BlueSeal~\cite{svt+14}). For instance, one of
our implemented operators for evaluating tools that detect data leaks is
as described in Listing~\ref{lst:operator_leak}.  
Moreover, security operators generalize to other security goals as well;
a simple operator for evaluating tools that detect vulnerable SSL use
(\eg MalloDroid) could add a \verb|TrustManager| with a vulnerable
\verb|isServerTrusted| method that returns true, which, when combined
with our expressive mutation schemes
(Section~\ref{sec:mutation-scheme}), would generate a diverse set of
mutants.

\begin{lstlisting}[basicstyle=\ttfamily\scriptsize,float,caption={{\small
  Security operator that injects a data leak from the Calendar API
  access to the device log.}},belowcaptionskip=-8mm,label=lst:operator_leak,emph={Inject},emphstyle=\bfseries]
Inject: 
  String dataLeak## = java.util.Calendar.getInstance().getTimeZone().getDisplayName();
  android.util.Log.d("leak-##", dataLeak##);
\end{lstlisting}

To derive security operators at the granularity of the security goal, we
must examine the claims made by existing tools; \ie security tools must
certainly detect the unwanted behavior that they claim to detect,
unless affected by some unsound design choice that hinders detection. In
order to precisely identify what a tool considers as a security flaw, and claims
to detect, we inspected the following sources:

\myparagraph{1) Research Papers} The tool's research paper is often the
primary source of information about what unwanted behavior a tool seeks
to detect. We inspect the properties and variations of the unwanted
behavior as described in the paper, as well as the examples provided, to
formulate security operator specifications
for injecting the unwanted behavior in an app. However, we do not
create operators using the limitations and assumptions already
documented in the paper or well-known in general (\eg leaks in reflection and dynamically loaded code), as \tool seeks to find unknown assumptions.

\myparagraph{2) Open source tool documentation} Due to space limitations
or tool evolution over time, research papers may not always have the
most complete or up-to-date information considering what security flaws
a tool can actually address. We used tool documentation available in
online appendices and open source repositories to fill this knowledge
gap. 

\myparagraph{3) Testing toolkits} Manually-curated testing toolkits (\eg
DroidBench~\cite{arf+14}) may be available, and may provide examples of
baseline operators.
    

%% file: mutation_schemes.tex
\vspace{-0.3em}
\subsection{Mutation Schemes}
\label{sec:mutation-scheme}
\vspace{-0.3em}
To enable the security evaluation of static analysis tools,
\tool must seed mutations within Android apps. We define the specific methods for
choosing \textit{where} to apply security operators to inject mutations
within Android apps as the mutation scheme. 

The mutation scheme depends on a number of factors: {\sf (1)}
Android's unique abstractions, {\sf (2)}, the intent to over-approximate
reachability for coverage, and {\sf (3)} the security goal of the tool
being analyzed (\ie similar to security operators).
Note that while
mutation schemes using the first two factors may be generally applied to
any type of static analysis tool (\eg SSL vulnerability and
malware detectors), the third factor, as the description suggests, will
only apply to a specific security goal, which in the light of this paper,
is data leak detection.

We describe each factor independently, as a mutation scheme, in the
context of the following running example described previously in
Section~\ref{sec:motivation}:

Recall that FlowDroid~\cite{arf+14}, the target of our analysis in
Section~\ref{sec:motivation}, detects data leaks in Android
apps. Hence, FlowDroid 
loosely defines a data leak as a flow
from a sensitive \textit{source} of information to some \textit{sink}
that exports it. FlowDroid lists all of the sources and sinks within a
configurable ``SourcesAndSinks.txt'' file in its tool documentation, from which it first selects a
simple source \verb|java.util.Calendar.getTimeZone()| and a simple sink
\verb|android.util.Log.d()|. We then design a
data leak operator, as shown in Listing~\ref{lst:operator_leak}.
Using this security operator, we implement the following three different mutation
schemes.

\subsubsection{Leveraging Android Abstractions} 
\label{sec:android-scheme}
The Android platform and
app model support numerous abstractions that pose
challenges to static analysis. One commonly stated example is the
absence of a {\sf Main} method as an entry-point into the app, which
compels static analysis tools to scan for the various entry points, and
treat them all similarly to a traditional {\sf Main} method~\cite{arf+14, Halavanalli:2013}.

Based on our domain knowledge of Android and its security, we choose
the following features as a starting point in a mutation scheme that
models unique aspects of Android, and  more importantly, tests the
ability of analysis tools to detect unwanted behavior
placed within these features (Goal~\ref{goal:android}):

\myparagraph{1) Activity and Fragment lifecycle}  Android apps
are organized into a number of {\em activity} components, which form the
user interface (UI) of the app. The activity lifecycle is controlled via
a set of callbacks, which are executed whenever an app is
launched, paused, closed, started, or stopped~\cite{android-lifecycle}. Fragments are also UI elements that possess similar
callbacks, though they are often used in a manner secondary to
activities.  We design our mutation scheme to place mutants
within methods of fragments and activities where applicable,
so as to test a tool's ability to model the activity and fragment
lifecycles.

\myparagraph{2) Callbacks}
Since  much of Android relies on
callbacks triggered by events, these callbacks pose a significant
challenge to traditional static analyses, as their code can be executed
asynchronously in several different potential orders. We place mutants within these
asynchronous callbacks to test the tools' ability to soundly model the
asynchronous nature of Android.  For instance, consider the example in
Listing~\ref{lst:onClick}, 
where the {\sf onClick()} callback can execute at any point of time.

\begin{lstlisting}[basicstyle=\ttfamily\scriptsize,float,caption={{\small
  Dynamically created {\sf onClick} callback}},belowcaptionskip=-8mm,label=lst:onClick,emph={},emphstyle=\bfseries]
final Button button = findViewById(R.id.button_id);
button.setOnClickListener(new View.OnClickListener() {public void onClick(View v) {// Code here executes on main thread after user presses button}});
\end{lstlisting}

\myparagraph{3) Intent messages} Android apps communicate with
one another and listen for system-level events using Intents, Intent
Filters, and Broadcast Receivers~\cite{android-intents,android-broadcasts}.
Specifically, Intent Filters and Broadcast Receivers form another major
set of callbacks into the app. Moreover, Broadcast Receivers can
be dynamically registered. Our mutation scheme not only places mutants
in the statically registered callbacks such as those triggered by
Intent Filters in the app's Android Manifest, but also callbacks
dynamically
registered within the program, and even within other callbacks, \ie
recursively. For instance, we generate a dynamically registered
broadcast receiver inside another dynamically registered broadcast
receiver, and instantiate the security operator within the inner
broadcast receiver (see Listing~\ref{lst:inception} in
Appendix~\ref{app:code} for the code). 

\myparagraph{4) XML resource files} Although Android apps are primarily
written in Java, they also include resource files that establish
callbacks. Such resource files also allow the developer to register for
callbacks from an action on a UI object (\eg the {\sf onClick} event,
for callbacks on a button being touched). As described previously,
static analysis tools often list these callbacks on par with the {\sf
Main} function, \ie as one of the many entry points into the app. We
incorporate these resource files into our mutation scheme, \ie mutate
them to call our specific callback methods.

\subsubsection{Evaluating Reachability} 
\label{sec:reachability-scheme}

The objective behind this simple, but important, mutation scheme is to
exercise the reachability analysis of the tool being evaluated. We
inject mutants (\eg data leaks from our example) at the start of every
method in the app. While the previous schemes add methods to the
app (\eg new callbacks), this scheme simply verifies if the app
successfully models the bare minimum.

\subsubsection{Leveraging the Security Goal}
\label{sec:goal-scheme}

Like security operators, mutation schemes may also be designed in a way
that accounts for the security goal of the tool being evaluated
(Goal~\ref{goal:operator}). Such schemes may be applied to any tool with
a similar objective. In keeping with our motivating example
(Section~\ref{sec:motivation}) and our evaluation
(Section~\ref{sec:eval}), we develop an example mutation scheme
that can be specifically applied to evaluate data leak detectors. This scheme infers two ways of adding mutants:

\myparagraph{1) Taint-based operator placement}
This placement methodology tests the tools' ability to recognize an
asynchronous ordering of callbacks, by placing {\em the source in one
callback and the sink in another}. The execution of
the source and sink may be triggered due to the user, and the app
developer (\ie especially a malicious adversary) may craft the mutation
scheme specifically so that the sources and sinks lie on callbacks that
generally execute in sequence. However, this sequence may not be
observable through just static analysis. A simple example is collecting
the source data in the {\sf onStart()} callback, and leaking it in the
{\sf onResume()} callback.
As per the activity lifecycle, the {\sf onResume()} callback {\em
always} executes right after the {\sf onStart()} callback.  

\myparagraph{2) Complex-Path operator placement}
Our preliminary analysis demonstrated that static analysis tools may
sometimes stop after an arbitrary number of hops when analyzing a call
graph, for performance reasons. This finding motivated the complex-path
operator placement. In this scheme, we make the path between source and
sink as complex as possible (\ie which is ordinarily one line of code,
as seen in Listing~\ref{lst:operator_leak}).
That is, the design of this scheme allows the injection of code along the path from source
to sink based on a set of predefined rules. In our evaluation, we
instantiate this scheme with a rule that recreates the String variable
saved by the source, by passing each
character of the string into a {\sf StringBuilder}, then sending the resulting
string to the sink. \tool allows the analyst to dynamically implement such
rules, as long as the input and output are both strings, and the rule
complicates the path between them by sending the input through
an arbitrary set of transformations.

In a traditional mutation analysis setting, the mutation placement
strategy would seek to minimize the number of non-compilable mutants. 
However, as our goal is to evaluate the soundness of Android
security tools, we design our mutation scheme to over-approximate. Once
the mutated apps are created, for a feasible analysis, we pass them
through a dynamic filter that removes the mutants that cannot be
executed, ensuring that the mutants that each security tool is evaluated against
are all executable.

%% file: analysis_feasibility.tex
\vspace{-0.2em}
\subsection{Analysis Feasibility \& Methodology}
\vspace{-0.3em}
\label{sec:feasibility}
\begin{figure}[t]
  \centering
  \includegraphics[width=1.4in]{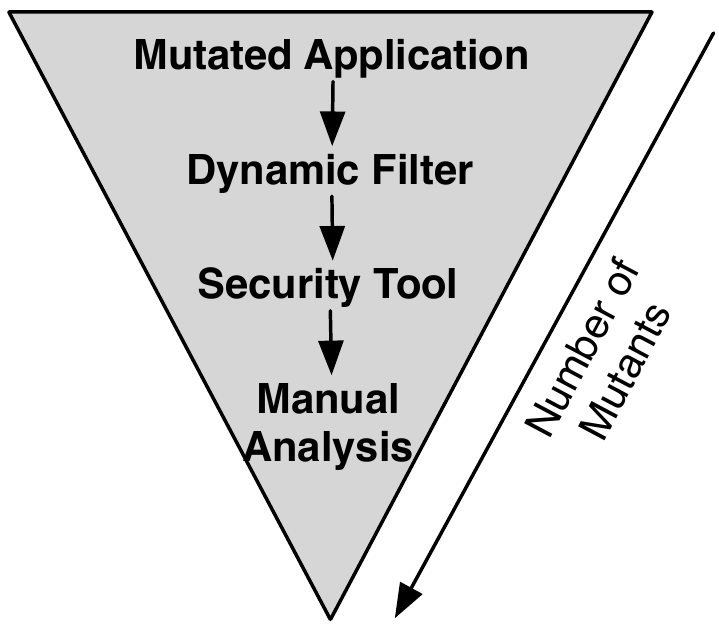}
  \vspace{-1em}
  \caption{{\small The number of mutants (\eg data leaks) to analyze
  drastically reduces at every stage in the process.}}
  \vspace{-1.5em}
\label{fig:filter}
\end{figure}

\tool reduces manual effort by filtering out mutants whose security
flaws are not verified by dynamic analysis
(Goal~\ref{goal:feasibility}). As described in Figure~\ref{fig:design},
for any given mutated app, we use a dynamic filter (\ie the Execution
Engine (EE), described in Section~\ref{sec:implementation}) to purge non-executable leaks.  If a mutant (\eg a data
leak) exists in the mutated app, but is not confirmed as executable by
the filter, we discard it. For example, data leaks injected in dead code
are filtered out. Thus, when the Android security tools are applied to
the mutated apps, only mutants that were executed by EE are considered.

Furthermore, after the security tools were applied to mutant apps,
only \textit{undetected} mutants are considered during analyst analysis.
The reduction in the number of mutants subject to analysis at each step
of the \tool process is illustrated in Figure~\ref{fig:filter}.

The following
methodology is used by an analyst for each undetected mutant 
after testing a given security tool to
isolate and confirm flaws:

    \myparagraph{1) Identifying the Source and Sink} During mutant
    generation, \tools ME injects a unique mutant
    identifier, as well as the source and sink using {\sf
    util.Log.d} statements. Thus, for each
    undetected mutant, an analyst simply looks up the unique IDs
    in the source to derive the source and sink.

	\myparagraph{2) Performing Leak Call-Chain Analysis} Since the
	data leaks under analysis went undetected by a given static
	analysis tool, this implies that there exists one (or multiple)
	method call sequences (\ie call-chains) invoking the source and
	sink that could not be modeled by the tool.  Thus, a security
	analyst inspects the code of a mutated app, and
	identifies the observable call sequences from various
	entry points.  This is aided by dynamic
	information from the EE so that an analyst can examine
	the order of execution of detected data leaks to infer the
	propagation of leaks through different call chains.

	\myparagraph{3) Synthesizing Minimal Examples} For each of the
	identified call sequences invoking a given undetected data
	leak's source and sink, an analyst then attempts to synthesize a
	minimal example by re-creating the call sequence using only the
	required Android APIs or method calls from the mutated app.
	This info is then inserted into a pre-defined skeleton
	app project so that it can be again analyzed by the
	security tools to confirm a flaw.

\myparagraph{4) Validating the Minimal Example} Once the minimal example
has been synthesized by the analyst, it must be validated against
the security tool that failed to detect it earlier.  If the tool fails
to detect the minimal example, then the process ends with the
confirmation of a flaw in the tool.  If the tool is
able to detect the examples, the analyst can either iteratively
refine the examples, or discard the mutant, and move on to the next
example.

%% file: implementation.tex
\section{Implementation}
\label{sec:implementation}

This section provides the implementation details of \tool: {\sf (1)} ME
for mutating apps, and {\sf (2)} EE for exercising mutants to filter out
non-executing ones. We have made \tool available for use by the wider
security research community~\cite{appendix}, along with the data
generated or used in our experiments (\eg operators, flaws) and code
samples.

\myparagraph{1. Mutation Engine (ME)} The ME allows \tool to
automatically mutate apps according to a fixed set of security operators
and mutation schemes. ME is implemented in Java and builds upon the {\sc
\small MDroid+} mutation framework for Android \cite{lbt+17}. Firstly,
ME derives a mutant injection profile (MIP) of all possible injection
points for a given mutation scheme, security operator, and target app
source code.  The MIP is derived through one of two types of analysis:
(i) text-based parsing and matching of \texttt{xml} files in the case of
app resources; or (ii) using Abstract Syntax Tree (AST) based analysis
for identifying potential injection points in code. \tool takes a
systematic approach toward applying mutants to a target app, and for
each mutant location stipulated by the MIP for a given app, a mutant is
seeded.  The injection process also uses either text- or AST-based code
transformation rules to modify the code or resource files.  In the
context of our evaluation, \tool further marks injected mutants in the
source code with log-based indicators that include a unique identifier
for each mutant, as well as the source and sink for the injected leak.
This information can be customized for future security operators and
exported as a ledger that tracks mutant data.  \tool can be extended to
additional security operators and mutation schemes by adding methods to
derive the MIP, and perform target code transformations.

Given the time cost in running the studied security-focused static
analysis tools on a set of \texttt{apks}, \tool breaks from the process
used by traditional mutation analysis frameworks that seed each mutant
into a separate program version, and seeds all mutants into a single
version of a target app.  Finally, the target app is automatically
compiled using its build system (\eg gradle~\cite{gradle},
ant \cite{ant}) so that it can be dynamically analyzed by the EE.

\myparagraph{2. Execution Engine (EE)}  To facilitate a feasible manual
analysis of the mutants that are undetected by a security analysis tool,
\tool uses the EE to dynamically analyze target apps, verifying whether
or not injected mutants can be executed in practice.  This EE builds
upon prior work in automated input generation for Android apps by
adapting the systematic exploration strategies from the {\sc \small
CrashScope} tool \cite{Moran:ICST16, Moran:ICSE17} to explore a target
app's GUI. We discuss the limitations of the EE in
Section~\ref{sec:limitations}. For more details, please see
Appendix~\ref{app:crashscope}.

%% file: evaluation.tex
\section{Evaluation}
\label{sec:eval}

The main \textit{goal} of our evaluation is to measure the effectiveness
of \tool at uncovering flaws in security-focused static
analysis tools for Android apps, and to demonstrate the extent of such
flaws. For this study, we focus on tools that detect private
data leaks on a device.  Specifically, we focus on a set of
\textit{seven} data leak detectors for Android that use static analysis,
primarily due to the availability of their source code, namely
FlowDroid~\cite{arf+14}, Argus~\cite{wror14} (previously known as
AmanDroid), DroidSafe~\cite{gkp+15}, IccTA~\cite{lbb+15},
BlueSeal~\cite{svt+14}, HornDroid~\cite{cgm16}, and
DidFail~\cite{kfb+14}. 
For all the tools except FlowDroid, we use the latest release
version when available; in FlowDroid's case, we used its v$2.0$ release for
our \tool analysis, and confirmed our findings with its later releases
(\ie v$2.5$ and v$2.5.1$). 
Additionally, we use a set of $7$ open-source Android apps from
F-droid~\cite{fdroid}~  
that we mutate. These $7$ apps produced 2026
mutants to inspect, which led to the discovery of $13$ flaws. A larger
dataset of apps is likely to generate more mutants, and lead to more
flaws.

In this section, we describe the highlights of our evaluation
(Section~\ref{sec:eval-highlights}), along with the \textit{three}
experiments we conduct, and their results.  In the first
experiment (Section~\ref{sec:executing_tools}), we run \tool on
\textit{three} tools, and record the number of leaks that each tool
fails to detect (\ie the number of uncaught mutants). In the second
experiment (Section~\ref{sec:flowdroid_vuln}), we perform an in-depth
analysis of FlowDroid by applying our systematic manual analysis
methodology (Section~\ref{sec:feasibility}) on the output of \tool for
FlowDroid. Finally, our third experiment
(Section~\ref{sec:vuln_propagation}) measures the propagation and
prevalence of the flaws found in FlowDroid, in tools from our
dataset apart from FlowDroid, and two newer versions of FlowDroid.

These experiments are motivated by the following research questions:

\begin{enumerate}[label=\textbf{RQ\arabic*},ref=\textbf{RQ\arabic*}]
\setlength{\itemsep}{-3pt}
  \item \label{rq:findings} {\em Can \tool find security
    problems in static analysis tools for Android, and help resolve them to
    flaws/ unsound choices?}
  \item \label{rq:inherit} {\em Are flaws inherited when a
    tool is reused (or built upon) by another tool?}
  \item \label{rq:feasible} {\em Does the semi-automated methodology of
    \tool allow a feasible analysis (in terms of manual effort)?}
  \item \label{rq:patch} {\em Are all flaws unearthed by \tool
    difficult to resolve, or can some be isolated and patched?}
  \item \label{rq:perf} {\em How robust is \tool's performance?}
\end{enumerate}

\subsection{Evaluation overview and Highlights}
\label{sec:eval-highlights}

We insert a total of $7,584$ data leaks (\ie mutants) in a set of $7$
applications using \tool.  $2,026$ mutants are verified as executable by
the EE, and $83$-$1,480$ are not detected depending on the studied tool.
During our analysis, \tool exhibits a maximum one-cost runtime of $92$
minutes (\ref{rq:perf}), apart from the time taken by the analyzed tool
(\eg FlowDroid) itself.  Further, our in-depth analysis of the output of
\tool for FlowDroid discovers $13$ unique flaws that are not
documented in either the paper or the source code repository
(\ref{rq:findings}). Moreover, it takes our analyst, a graduate student
with background in Android security, {\em one} hour per flaw
(in the worst case), due to our systematic analysis methodology, as well
as our dynamic filter (Section~\ref{sec:feasibility}), which filters out
over $73$ \% of the seeded non-executable mutants (\ref{rq:feasible}).
Further, we demonstrate that two newer versions of FlowDroid, as well as
the {\em six} other tools set apart from FlowDroid (including those that
inherit it), are also vulnerable to at least {\em one} flaw
detected in FlowDroid (\ref{rq:inherit}).  This is confirmed, with
negligible effort, using minimal examples generated during our analysis
of FlowDroid (\ref{rq:feasible}).  Finally, we are able to generate
patches for a specific flaw discovered in FlowDroid, and our
pull request has been accepted by the tool authors (\ref{rq:patch}).

\input{executing_tools}
\input{flowdroid_vuln}

\input{vuln_propagation}

%% file: executing_tools.tex
\subsection{Executing \tool}
\label{sec:executing_tools}

The objective of this experiment is to demonstrate the effectiveness of
\tool in filtering out non-executable injected leaks (\ie mutants),
while illustrating that this process results in a reasonable number of
leaks for an analyst to manually examine.

\myparagraph{Methodology}
We create $21$ mutated APKs from $7$ target applications, with $7,584$
leaks among them, by combining the security operators described in
Section~\ref{sec:operators}, with mutation schemes from
Section~\ref{sec:mutation-scheme}.  First, we measure the total number
of leaks injected across all apps, and then the total number of leaks
marked by the EE as non-executable.  Note that this number is
independent of the tools involved, \ie the filtering only happens once,
and the mutated APKs can then be passed to any number of tools for
analysis. The non-executable leaks are then removed.  Next, we configure
FlowDroid, Argus, and DroidSafe and evaluate each tool with \tool
individually, by running them on the mutated apps (with non-executable
leaks excluded) and recording the number of leaks not detected by each
tool (\ie the {\em surviving} mutants).

\input{tables/tool_table}

\myparagraph{Results} \tool injects $7,584$ leaks into the Android apps,
of which, $5,558$ potentially non-executable leaks are filtered out using
our EE, leaving only $2,026$  leaks confirmed as executable in
the mutated apps.  By filtering out a large number of potentially
non-executable leaks (\ie over 73\%), our dynamic filtering significantly
reduces manual effort (\ref{rq:feasible}).

Table~\ref{tbl:tool_table} shows the statistics acquired from \tool's
output over FlowDroid, Argus, and DroidSafe. We observe that
FlowDroid cannot detect over 48\% of the leaks, while Argus cannot
detect over 73\%. Further, DroidSafe does not detect a non-negligible
percentage of leaks (\ie over 4\%), and as these leaks have been
confirmed to execute  by our EE, it is safe to say that DroidSafe has
flaws as well.  
Note that this experimental result validates our conceptual
argument, that security operators designed for a specific goal may apply
to tools with that goal. 
However, given its popularity, we limit our
in-depth evaluation to FlowDroid.

Finally, we measure the runtime of the \tool-specific part of the
analysis, \ie up to executing the tool to be evaluated, to be a constant
$92$ minutes in the worst case, a majority of which (\ie $99$\%) is
taken up by the EE. Note that the time taken by \tool is a one-time
cost, and does not have to be repeated for tools with a similar security
goal (\ref{rq:perf}).

%% file: tables/tool_table.tex
\begin{table}
\scriptsize
\centering
\caption{{\small The number and percentage of leaks not detected by $3$
popular data leak detection tools.  }}
\label{tbl:tool_table}
\setlength{\tabcolsep}{2pt}
\begin{tabular}{c|c|c}
  \Xhline{2\arrayrulewidth}
  {\bf Tool}	& {\bf Undetected Leaks} & {\bf
  Undetected Leaks (\%)} \\
  \Xhline{2\arrayrulewidth}
  FlowDroid v2.0 & 987$/$2,026  & 48.7\% \\
  Argus & 1,480$/$2,026  &  73.1\%\\
  DroidSafe & 83$/$2,026  & 4.1\% \\
  \Xhline{2\arrayrulewidth}
\end{tabular} 
\vspace{-2em}
\end{table}

%% file: flowdroid_vuln.tex
\subsection{FlowDroid Analysis}
\label{sec:flowdroid_vuln}
\input{tables/vuln_table}

This experiment demonstrates an in-depth, manual analysis of FlowDroid,
which we choose for two
reasons: {\sf (1)} impact (FlowDroid is cited by 700 papers and numerous
other tools depend on it), and {\sf (2)} potential for change (since FlowDroid is being maintained at the moment, any contributions we
can make will have immediate benefits).

\myparagraph{Methodology} We performed an in-depth analysis using the
list of surviving mutants (\ie undetected leaks) generated by \tool for
FlowDroid v$2.0$ in the previous experiment. We leveraged the
methodology for systematic manual evaluation, described in
Section~\ref{sec:feasibility}, and discovered $13$ unique
flaws. {\em We confirmed that none of the discovered
flaws have been documented before}; \ie in the FlowDroid paper
or in their official documentation.

\myparagraph{Results} We discovered $13$ unique flaws, from FlowDroid
alone, demonstrating that \tool can be effectively used to find problems
that can be resolved to flaws (\ref{rq:findings}).  Using the approach
from Section~\ref{sec:feasibility}, the analyst  
needed less than an hour to isolate a flaw from the set of undetected
mutants, in the worst case.  In the best case, flaws were found in a
matter of minutes, demonstrating that the amount of manual effort
required to quickly find flaws using \tool is minimal
(\ref{rq:feasible}).  We give descriptions of the flaws discovered as a
result of \tools analysis in Table \ref{tab:flowdroid-vuln-descs}.

We have reported these flaws, and are working with the
developers to resolve the issues. In fact, we developed patches to
correctly implement Fragment support (\ie flaw $13$ in
Table~\ref{tab:flowdroid-vuln-descs}), which were accepted 
by developers.

To gain insight about the practical challenges faced by static analysis
tools, and their design flaws, we further categorize the discovered
flaws into the following {\em flaw classes}: 

\myparagraph{FC1: Missing Callbacks} The security tool (\eg FlowDroid)
does not recognize some callback method(s), and will not find leaks
placed within them. Tools that use lists of APIs or callbacks are
susceptible to this problem, as prior work has demonstrated as the
generated list of callbacks {\sf (1)} may not be complete, and {\sf (2)}
or may not be updated as the Android platform evolves.  
We found both these cases in our analysis of FlowDroid. That is, {\sf
DialogFragments} was added in API 11, \ie {\em before FlowDroid was
released}, and {\sf NavigationView} was added after.  These limitations
are well-known in the community of researchers at the intersection of
program analysis and Android security, and have been documented by prior
work~\cite{cfb+15}.  However, \tool helps evaluate the robustness of
existing security tools against these flaws, and helps in uncovering
these undocumented flaws for the wider security audience.  Additionally,
{\em some of these flaws may not be resolved even after adding the
callback to the list}; \eg {\sf PhoneStateListener} and {\sf
SQLiteOpenHelper}, both added in API 1, are not interfaces, but abstract
classes. Therefore, adding them to FlowDroid's list of callbacks (\ie
{\sf AndroidCallbacks.txt}) does not resolve the issue.

\myparagraph{FC2: Missing Implicit Call} The security tool does not
identify leaks within some method that is implicitly called by another
method.  For instance, FlowDroid does not recognize the path to {\sf
Runnable.run()} when a Runnable is passed into the {\sf
ExecutorService.submit(Runnable)}. The response from the developers
indicated that this class of flaws was due to an unresolved design
challenge in Soot's~\cite{vcg+99} SPARK algorithm, upon which FlowDroid
depends. 
This limitation is also known within the program analysis
community~\cite{cfb+15}. However, the documentation of this gap, thanks
to \tool, would certainly help developers and researchers in the wider
security community.

\myparagraph{FC3: Incorrect Modeling of Anonymous Classes} The security tool
misses data leaks expressed within an anonymous class. For example,
FlowDroid does not recognize leaks in the {\sf onReceive()} callback of a
dynamically registered {\sf BroadcastReceiver}, which is implemented within
another dynamically registered BroadcastReceiver's {\sf onReceive()} callback.
It is important to note that finding such complex flaws is
only possible due to \tool's semi-automated mechanism, and may be
rather prohibitive for an entirely manual analysis.

\myparagraph{FC4: Incorrect Modeling of Asynchronous Methods} The
security tool does not recognize a data leak whose source and sink are
called within different methods that are asynchronously executed. For
instance, FlowDroid does not recognize the flow between data leaks in
two callbacks (\ie\ {\sf onLocationChanged} and {\sf onStatusChanged})
of the {\sf LocationListener} class, which the adversary may cause to
execute sequentially (\ie as our EE confirmed).

Apart from {\bf FC1}, which may be patched with limited efforts, the
other three categories of flaws may require a significant amount of
research effort to resolve. 
However, documenting them is critical to
increase awareness of real challenges faced by Android static analysis
tools.

%% file: tables/vuln_table.tex
\begin{table*}[t]
\centering
\scriptsize
\caption{{\small Descriptions of flaws uncovered in FlowDroid
v$2.0$}}
\label{tab:flowdroid-vuln-descs}
\def\arraystretch{1.2}
\begin{tabular}{p{4cm}|p{11.7cm}}
\Xhline{2\arrayrulewidth}
\multicolumn{0}{c|}{\textbf{Flaw}}   & \multicolumn{1}{c}{\textbf{Description}}                                                                                                                                                                                                                                                                                                                                                                                                \\ \Xhline{2\arrayrulewidth}
\multicolumn{2}{c}{\textbf{FC1: Missing Callbacks}} \\ \Xhline{2\arrayrulewidth}

1. DialogFragmentShow          &  FlowDroid misses the {\sf DialogFragment.onCreateDialog()} callback registered by {\sf DialogFragment.show()}.\\ \hline

2. PhoneStateListener          & FlowDroid does not recognize the {\sf onDataConnectionStateChanged()} callback for classes extending the {\sf PhoneStateListener} abstract class from the telephony package.                                                                                                                                                                                                                                           \\ \hline

3. NavigationView            & FlowDroid does not recognize the {\sf onNavigationItemSelected()} callback of classes implementing the interface {\sf NavigationView.OnNavigationItemSelectedListener}.                                                                                                                                                                                                                                                         \\ \hline

4. SQLiteOpenHelper           & FlowDroid misses the {\sf onCreate()} callback of classes extending {\sf android.database.sqlite.SQLiteOpenHelper}.                                                                                                                                                                                                                                                                          \\ \hline

5. Fragments                 & FlowDroid 2.0 does not model Android
Fragments correctly. We added a patch, which was promptly accepted.
However, FlowDroid 2.5 and 2.5.1 remain affected. We investigate this further in the next section.                                                                                                                                                                                                 \\ \Xhline{2\arrayrulewidth}

\multicolumn{2}{c}{\textbf{FC2: Missing Implicit Calls}} \\ \Xhline{2\arrayrulewidth}

6. RunOnUIThread               & FlowDroid misses the path to {\sf Runnable.run()} for Runnables passed into {\sf Activity.runOnUIThread()}.                                                                                                                                                                                                                                                                                                              \\ \hline

7. ExecutorService             & FlowDroid misses the path to {\sf Runnable.run()} for Runnables passed into {\sf ExecutorService.submit()}.                                                                                                                                                                                                                                                                     \\ \Xhline{2\arrayrulewidth}

\multicolumn{2}{c}{\textbf{FC3: Incorrect Modeling of Anonymous Classes}} \\ \Xhline{2\arrayrulewidth}

8. ButtonOnClickToDialogOnClick & FlowDroid does not recognize the {\sf onClick()} callback of {\sf DialogInterface.OnClickListener} when instantiated within a Button's {\sf onClick=``method\_name''} callback defined in XML. FlowDroid will recognize this callback if the class is instantiated elsewhere, such as within an Activity's {\sf onCreate()} method.  \\ \hline

9. BroadcastReceiver           & FlowDroid misses the {\sf onReceive()} callback of a BroadcastReceiver implemented programmatically and registered within another programmatically defined and registered BroadcastReceiver's {\sf onReceive()} callback.    \\ \Xhline{2\arrayrulewidth}

\multicolumn{2}{c}{\textbf{FC4: Incorrect Modeling of Asynchronous Methods}} \\ \Xhline{2\arrayrulewidth}

10. LocationListenerTaint       & FlowDroid misses the flow from a source in the {\sf onStatusChanged()} callback to a sink in the {\sf onLocationChanged()} callback of the {\sf LocationListener} interface, despite recognizing leaks wholly contained in either.                                                                                                                                                                                                                                             \\ \hline
11. NSDManager                  & FlowDroid misses the flow from sources in any callback of a {\sf NsdManager.DiscoveryListener} to a sink in any callback of a {\sf NsdManager.ResolveListener}, when the latter is created within one of the former's callbacks.                                                                                                                                                                                                                         \\ \hline
12. ListViewCallbackSequential &  FlowDroid misses the flow from a source to a sink within different methods of a class obtained via {\sf AdapterView.getItemAtPosition()} within the {\sf onItemClick()} callback of an {\sf AdapterView.OnItemClickListener}.  \\ \hline
13. ThreadTaint                & FlowDroid misses the flow to a sink within a {\sf Runnable.run()} method started by a Thread, only when that Thread is saved to a variable before {\sf Thread.start()} is called. \\ \Xhline{2\arrayrulewidth}                                                                                                                                        
\end{tabular}
\vspace{-2.5em}
\end{table*}

%% file: vuln_propagation.tex
\subsection{Flaw Propagation Study}
\label{sec:vuln_propagation}

\input{tables/tool_vuln_table}

The objective of this experiment is to determine if the flaws
discovered in FlowDroid have propagated to the tools that inherit it, and
to determine whether other static analysis tools that do not
inherit FlowDroid are similarly flawed.

\myparagraph{Methodology} We check if the two newer release
versions of FlowDroid (\ie v$2.5$, and v$2.5.1$), as well as $6$ other
tools (\ie Argus, DroidSafe, IccTA, BlueSeal, HornDroid, and DidFail),
are susceptible to any of the flaws discussed previously in
FlowDroid v$2.0$, by using the tools to analyze the minimal example APKs
generated during the in-depth analysis of FlowDroid.

\myparagraph{Results} 
As seen in the Table~\ref{tbl:vuln_table}, all the versions of FlowDroid
are susceptible to the flaws discovered from our analysis of
FlowDroid v2.0. Note that while we fixed the {\sf Fragment} flaw and our
patch was accepted to FlowDroid's codebase, the latest releases of
FlowDroid (\ie v2.5 and v2.5.1) still seem to have this flaw. We are
working with the developers on a solution.

A significant observation from the Table~\ref{tbl:vuln_table} is that
the tools that directly inherit FlowDroid (\ie IccTA, DidFail) are similarly
flawed as FlowDroid. This is especially true when the tools do not
augment FlowDroid in any manner, and use it as a black box
(\ref{rq:inherit}). On the contrary, Argus, which is motivated by
FlowDroid's design, but augments it on its own, does not exhibit as many
flaws.

Also, BlueSeal, HornDroid, and DroidSafe use a significantly
different methodology, and are also not susceptible to these
flaws. Interestingly, BlueSeal and DroidSafe are similar to
FlowDroid in that they use Soot to construct a control flow graph, and
rely on it to identify paths between sources and sinks. However,
BlueSeal and DroidSafe both augment the graph in novel ways, and thus
don't exhibit the flaws found in FlowDroid.

Finally, our analysis does not imply that FlowDroid is weaker than the
tools which have fewer flaws in Table~\ref{tbl:vuln_table}.
However, it does indicate that the flaws discovered may be
typical of the design choices made in FlowDroid, and inherited by the tools
such as IccTA and DidFail. A similar deep exploration into the results of \tool
for the other tools may be explored in the future
(\eg of the $83$ uncaught leaks in DroidSafe from
Section~\ref{sec:executing_tools}).

%% file: tables/tool_vuln_table.tex
\begin{table*}
\scriptsize
\centering
\caption{{\small Flaws present in data leak detectors. Note that
a ``$-$'' indicates tool crash with the minimal APK, a ``\checkmark''
indicates presence of the flaw, and a ``{\sf x}'' indicates absence,
and *{\sf FD} = {\sf FlowDroid.}}}
\label{tbl:vuln_table}
\setlength{\tabcolsep}{5pt}
\begin{tabular}{l|c|c|c|c|c|c|c|c|c}
  \Xhline{2\arrayrulewidth}
  {\bf Flaw}	& {\bf FD$^*$ v2.5.1} & {\bf FD$^*$ v2.5} & {\bf
  FD$^*$ v2.0} & {\bf Blueseal} & {\bf IccTA} & {\bf HornDroid} & {\bf Argus} & {\bf DroidSafe} & {\bf DidFail} \\
  \Xhline{2\arrayrulewidth}
  DialogFragmentShow		& \checkmark & \checkmark & \checkmark & {\sf x} & \checkmark & \checkmark & {\sf x} & {\sf x} & \checkmark\\

  PhoneStateListener		& \checkmark & \checkmark & \checkmark & {\sf x} & \checkmark & \checkmark & {\sf x} & {\sf x} & \checkmark\\

  NavigationView		& \checkmark & \checkmark & \checkmark & - & \checkmark & - & \checkmark & - & \checkmark\\

  SQLiteOpenHelper		& \checkmark & \checkmark & \checkmark & {\sf x} & \checkmark & \checkmark & \checkmark & {\sf x} & \checkmark\\

  Fragments			& \checkmark & \checkmark & \checkmark & \checkmark & \checkmark & \checkmark & \checkmark & - & \checkmark\\

  RunOnUIThread			& \checkmark & \checkmark & \checkmark & {\sf x} & \checkmark & \checkmark & \checkmark & {\sf x} & \checkmark\\

  ExecutorService		& \checkmark & \checkmark & \checkmark & {\sf x} & \checkmark & \checkmark & \checkmark & {\sf x} & \checkmark\\

  ButtonOnClickToDialogOnClick	& \checkmark & \checkmark & \checkmark & {\sf x} & \checkmark & {\sf x} & {\sf x} & \checkmark & \checkmark\\

  BroadcastReceiver		& \checkmark & \checkmark & \checkmark & {\sf x} & \checkmark & {\sf x} & {\sf x} & {\sf x} & \checkmark\\

  LocationListenerTaint		& \checkmark & \checkmark & \checkmark & {\sf x} & \checkmark & {\sf x} & {\sf x} & {\sf x} & \checkmark\\

  NSDManager			& \checkmark & \checkmark & \checkmark & {\sf x} & \checkmark & {\sf x} & \checkmark & {\sf x} & \checkmark\\

  ListViewCallbackSequential	& \checkmark & \checkmark & \checkmark & {\sf x} & \checkmark & {\sf x} & {\sf x} & {\sf x} & \checkmark\\

  ThreadTaint			& \checkmark & \checkmark & \checkmark & {\sf x} & \checkmark & {\sf x} & {\sf x} & {\sf x} & \checkmark\\
  
  \Xhline{2\arrayrulewidth}
\end{tabular} 
\vspace{-2em}
\end{table*}

%% file: discussion.tex
\section{Discussion}
\label{sec:discussion}

\tool has demonstrated efficiency and effectiveness at revealing real
undocumented flaws in prominent Android security analysis tools. 
While experts in Android static analysis may be familiar with some of
the flaws we discovered (\eg some flaws in FC1 and FC2), we aim to
document these flaws for the entire scientific community. Further, \tool
indeed found some design gaps that were surprising to expert developers;
\eg FlowDroid's design does not consider callbacks in anonymous inner
classes (flaws 8-9, Table~\ref{tbl:vuln_table}), and in our interaction
with the developers of FlowDroid, they acknowledged handling such
classes as a non-trivial problem.
During our evaluation of \tool we were able to glean the following
pertinent insights:

\myparagraph{Insight 1} {\em Simple and security goal-specific mutation schemes are
effective.} While certain mutation schemes may be Android-specific, our
results demonstrate limited dependence on these configurations. Out of
the 13 flaws discovered by \tool, the Android-influenced mutation scheme
(Section~\ref{sec:android-scheme}) revealed one (\ie\ {\em
BroadCastReceiver} in Table~\ref{tbl:vuln_table}), while the rest were
evenly distributed among the other two mutation schemes; \ie the schemes
that evaluate reachability (Section~\ref{sec:reachability-scheme}) or
leverage the security goal (Section~\ref{sec:goal-scheme}).

\myparagraph{Insight 2} {\em Security-focused static analysis tools exhibit
undocumented flaws that require further evaluation and
analysis}.  Our results clearly demonstrate that previously unknown
security flaws or undocumented design assumptions, which can
be detected by \tool, pervade existing Android security static analysis
tools. 
Our findings not only motivate the dire need for systematic discovery,
fixing and documentation of unsound choices in these tools, but also
clearly illustrate the power of mutation based analysis adapted in
security context. 

\myparagraph{Insight 3} {\em Current tools inherit flaws from
legacy tools}. A key insight from our work is that while inheriting code
of the foundational tools (\eg FlowDroid) is a common practice, some of
the researchers may not necessarily be aware of the unsound choices they
are inheriting as well.  As our study results demonstrate, when a tool
inherits another tool directly (\eg IccTA inherits FlowDroid), all the
flaws propagate. More importantly, even in those cases where
the tool does not directly inherit the codebase, unsound choices may
still propagate at the conceptual level and result in real flaws.

\myparagraph{Insight 4} {\em As tools, libraries, and the Android
platform evolve, security problems become harder to track down}. Due the
nature of software evolution, all the analysis tools, underlying
libraries, and the Android platform itself evolve asynchronously. A few
changes in the Android API may introduce undocumented flaws in analysis
tools. \tool handles this fundamental obstacle of
continuous change by ensuring that each version of an analysis tool is
systematically tested, as we realize while tracking the Fragment flaw in
multiple versions of FlowDroid. 

\myparagraph{Insight 5} {\em Benchmarks need to evolve with time}.  While
manually-curated benchmarks (\eg DroidBench \cite{arf+14}) are highly
useful as a "first line of defense" in checking if a tool is able to
detect well-known flaws, the downside of relying too heavily
on benchmarks is that they only provide a known, finite number of tests,
leading to a false sense of security. Due to constant changes (insight
\#3) benchmarks are likely to become less relevant unless they are
constantly augmented, which requires tremendous effort and coordination.
\tool significantly reduces this burden on benchmark creators via its
suite of extensible and expressive security operators and mutation
schemes, which can continuously evaluate new versions of tools. The key
insight we derive from our experience building \tool is that {\em while
benchmarks may check for documented flaws, \tool's true
strength is in discovering new flaws.}

%% file: related-work.tex
\section{Related Work}
\label{sec:relwork}

\tool builds upon the theoretical underpinnings of mutation analysis
from SE, and to our knowledge, is the first work to adapt mutation
analysis to evaluate the soundness claimed by security tools. Moreover,
\tool adapts mutation analysis to security, and makes fundamental and
novel modifications (described previously in Section~\ref{sec:design}).
In this section, we survey related work in three other related areas:

\myparagraph{Formally Verifying Soundness} While an ideal approach, 
formal verification is one of the most difficult problems in computer
security.  For instance, prior work on formally verifying apps often
requires the monitor to be rewritten in a new language or use
verification-specific programming constructs (\eg verifying reference
monitors~\cite{fcds10,vcj+13}, information flows in
apps~\cite{mye99,ml00,yys12}), which poses practical concerns for tools
based on numerous legacy codebases (\eg FlowDroid~\cite{arf+14},
CHEX~\cite{llw+12}). Further, verification techniques generally require
correctness to be specified, \ie the policies or invariants that the
program is checked against. Concretely defining what is ``correct'' is
hard even for high-level program behavior (\eg making a ``correct'' SSL
connection), and may be infeasible for complex static analysis tools
(\eg detecting ``all incorrect SSL connections'').  \tool does not aim
to substitute formal verification of static analysis tools; instead, it
aims to uncover existing limitations of such tools.

\myparagraph{Mutation Analysis for Android}
 Deng \etal \cite{Deng:ICSTW15} introduced mutation analysis for Android and derived
 operators by analyzing the syntax of Android-specific Java constructs.  
Subsequently,  a mutation analysis framework for Android ($\mu$Droid)
has been introduced to evaluate a test suite's
ability to uncover energy bugs \cite{Jabbarvand:FSE17}.  
\tool incorporates concepts from the general mutation analysis proposed
by prior work (especially on
Android~\cite{Jabbarvand:FSE17,Deng:ICSTW15,lbt+17}), but adapts them in
the context of security. We design mSE to focus on undetected mutants,
providing a semi-automated methodology to resolve  such mutants to
design/implementation flaws (Section~\ref{sec:feasibility}). The
derivation of security operators (Section~\ref{sec:operators})
represents a notable departure from traditional mutation testing that
seeds simple syntactic code changes.  Our mutation schemes
(Section~\ref{sec:mutation-scheme}) evaluate coverage of OS-specific
abstractions, reachability of the analysis, or the ability to detect
semantically-complex mutants, providing the expressibility necessary for
security testing, while building upon traditional approaches.  
Further, \tool builds upon the software infrastructure developed for
{\sc \small MDroid+}~\cite{lbt+17} that allows a scalable analysis of
mutants seeded according to security operators. In particular, \tool
adapts the process of deriving a potential fault profile for mutant
injection and relies on the EE to validate the mutants seeded according
to our derived security operators. 

\myparagraph{Android Application Security Tools}  
The popularity and open-source nature of Android has spurred an
immense amount of research related to examining and improving the
security of the underlying OS, SDK, and apps.
Recently, Acar \etal have systematized  
Android security research \cite{androidsok:sp16}, and we discuss work that
introduces static analysis-based countermeasures for Android security
issues according to Acar \etals categorization. 

Perhaps the most prevalent area of research in Android security has
concerned the permissions system that mediates access to privileged
hardware and software resources.  Several approaches have motivated
changes to Android's permission model, or have proposed enhancements to
it, with goals ranging from detecting or fixing unauthorized information
disclosure or leaks in third party
applications~\cite{egc+10,arf+14,gcec12,ne13,naej16,xw15,jaf+13} to 
detecting overprivilege in applications~\cite{fch+11,azhl12,vcc11}.
Similarly, prior work has also focused on benign but vulnerable Android
applications, and proposed techniques to detect or fix vulnerabilities
such as cryptographic API misuse API~\cite{fhm+12,ebfk13,ssl+14,fhp+13}
or unprotected application interfaces~\cite{fwm+11,cfgw11,lbs+17}.
Moreover, these techniques have often been deployed as modifications to
Android's permission
enforcement~\cite{eom09b,egc+10,nkz10,Fragkaki:ESORICS12,dsp+11,fwm+11,bdd+12,omem09,cnc10,zzjf11,sc13,bdd+11b,hnes14,pfnw12,sdw12,rk13},
SDK tools~\cite{fch+11,azhl12,vcc11}, or inline reference
monitors~\cite{xsa12,jmv+12,Davis2012IARMDroidAR,bgh+13,bbh+15}.
While this paper demonstrates the evaluation of only a small subset of
these tools with \tool, our experiments demonstrate that \tool has the
potential to impact nearly all of them.  For instance, \tool could be
applied to further vet SSL analysis tools by purposely introducing
complex SSL errors in real applications, or tools that analyze
overprivilege or permission misuse, by developing security operators
that attempt to misuse permissions to circumvent such monitors. Future
work may use \tool to perform an in-depth analysis of these problems.

%% file: limitations.tex
\section{Limitations}
\label{sec:limitations}

\myparagraph{1) Soundness of \tool} As acknowledged in
Section~\ref{sec:relwork}, mSE does not aim to supplant formal
verification (which would be sound), and does not claim soundness
guarantees. Rather, mSE provides a systematic approach to
semi-automatically uncover flaws in existing security tools, which is a significant
advancement over manually-curated tests.

\myparagraph{2) Manual Effort} Presently, the workflow of \tool requires
an analyst to manually analyze the result of \tool (\ie uncaught
mutants). However, as described in Section~\ref{sec:executing_tools},
\tool possesses enhancements that mitigate the manual effort by
dynamically eliminating non-executable mutants, that would otherwise
impose a burden on the analyst examining undetected mutants. In our
experience, this analysis was completed in a reasonable time using the
methodology outlined in Section~\ref{sec:feasibility}.

\myparagraph{3) Limitations of Execution Engine} Like any dynamic
analysis tool, 
the EE will not explore all
possible program states, thus, there may be a set of mutants marked as
non-executable by the EE, that may actually be executable
under certain scenarios.  However, the {\sc \small CrashScope} tool,
which \tools's EE is based upon, has been shown to
perform comparably to other tools in terms of
coverage \cite{Moran:ICST16}.  
Future versions of \tool's EE could rely on emerging
input generation tools for Android apps \cite{Mao:ISSTA16}.

%% file: conclusion.tex
\section{Conclusion}
\label{sec:conc}

We proposed the \tool framework for performing systematic
security evaluation of Android static analysis tools to discover (undocumented) unsound
assumptions, adopting the practice of mutation testing from SE to security.
\tool not only detected major flaws in a
popular, open-source Android security tool, but also demonstrated how
these flaws propagated to other tools that inherited the security
tool in question. With \tool, we demonstrated how mutation analysis can
be feasibly used for gleaning unsound assumptions in existing tools,
benefiting developers, researchers, and end users, by making
such tools more secure and transparent.

%% file: acknowledgements.tex
\section{Acknowledgements}

We thank Rozda Askari for his help with setting up experiments.  
We thank the FlowDroid developers, as well as the developers of the
other tools we evaluate in this paper, for making their tools available
to the community, providing us with the necessary information for our
analysis, and being open to suggestions and improvements.  The authors
have been supported in part by the NSF-1815336, NSF-714581 and
NSF-714161 grants.  Any opinions, findings, and conclusions expressed
herein are the authors' and do not necessarily reflect those of
the sponsors.

%% file: appendix.tex
\appendix

\section{Fragment Use Study}

We performed a small-scale app study using
the Soot \cite{soot} static analysis library to deduce how commonly
fragments were used in
real apps. That is, we analyzed 240 top apps from every
category on Google Play (\ie a total of 8,664 apps collected as of June
2017 after removing duplicates), and 
observed that at least 4,273 apps (49.3\%) used
fragments in their main application code, while an additional 3,587
(41.4\%) used fragments in packaged libraries. Note that while we
did not execute the apps to determine if the fragment code was really
executed, the fact that 7,860 out of 8,664 top apps, or ~91\% of
popular apps contain fragment code indicates the possibility
that fragments are widely used, and that accidental or malicious data
leaks in a large number of apps could evade FlowDroid due to
this flaw. 

\section{Code Snippets}
\label{app:code}
In Listing~\ref{lst:inception}, we dynamically register a broadcast receiver inside another dynamically registered broadcast receiver, and add the mutant (\ie a data leak in this case) inside the {\sf onReceive()} callback of the inner broadcast receiver.

\begin{lstlisting}[basicstyle=\ttfamily\scriptsize,float,caption={\small Dynamically created Broadcast Receiver, created inside another, with data leak.},belowcaptionskip=-8mm,label=lst:inception,emph={},emphstyle=\bfseries]
BroadcastReceiver receiver = new BroadcastReceiver() {
    @Override
    public void onReceive(Context context, Intent intent) {
    BroadcastReceiver receiver = new BroadcastReceiver(){
        @Override
            public void onReceive(Context context, Intent intent) {
                String dataLeak = Calendar.getInstance().getTimeZone().getDisplayName();
                Log.d("leak-1", dataLeak);
            }
    };
    IntentFilter filter = new IntentFilter();
    filter.addAction("android.intent.action.SEND");
    registerReceiver(receiver, filter);
}};
IntentFilter filter = new IntentFilter();
filter.addAction("android.intent.action.SEND");
registerReceiver(receiver, filter);     
\end{lstlisting}

\section{CrashScope (Execution Engine)}
\label{app:crashscope}
The EE functions by statically analyzing the code of a
target app to identify activities implementing potential
contextual features (\eg rotation, sensor usage) via API call-chain
propagation.  It then executes an app according to one of several
exploration strategies while constructing a dynamic event-flow model of
an app in an online fashion.  These strategies are organized along three
dimensions: (i) GUI-exploration, (ii) text-entry, and (iii) contextual
features.  The Execution Engine uses a Depth-First Search (DFS)
heuristic to systematically explore the GUI, either starting from the 
top of the screen down, or from the bottom of the screen up.  It is also
able to dynamically infer the allowable text characters from the Android
software keyboard and enter expected text or no text.  
Finally, the EE can exercise contextual features (\eg
rotation, simulating GPS coordinates).
Since the goal of the EE is to explore as many screens of a target app as possible, the 
EE forgoes certain combinations of exploration strategies
from {\sc \small CrashScope} (\eg entering unexpected text or disabling
contextual features) prone to eliciting crashes from apps.  The approach
utilizes \texttt{adb} and Android's 
\texttt{uiautomator} framework to interact with and extract GUI-related
information from a target device or emulator.